\documentclass[final,3p,11pt]{elsarticle}

\usepackage[utf8]{inputenc} 
\usepackage[T1]{fontenc}    
\usepackage{url}            
\usepackage{booktabs}       
\usepackage{amsmath}        
\usepackage{amssymb}        
\usepackage{nicefrac}       
\usepackage{microtype}      
\usepackage[linesnumbered,ruled,vlined]{algorithm2e}
\usepackage{listings}
\usepackage{xcolor}
\usepackage{enumitem}
\usepackage{caption}
\usepackage{float}
\usepackage{placeins}

\graphicspath{{media/}{images/}}     

\SetKwBlock{Begin}{begin}{end}
\SetKwFor{ForEach}{for each}{do}{endfor}
\SetKwIF{If}{ElseIf}{Else}{if}{then}{else if}{else}{endif}
\SetKwFor{While}{while}{do}{endwhile}
\SetKwFor{For}{for}{do}{endfor}
\SetKw{KwTo}{to}
\SetKw{Continue}{continue}
\SetKw{Break}{break}
\SetKw{Return}{return}
\SetKw{Downto}{downto}

\makeatletter
\let\ps@pprintTitle\ps@plain
\newcommand{\preprintauthors}{}
\newcommand{\preprintauthor}[5][]{%
  \author[#1]{#2}\ead{#3}%
  \g@addto@macro\preprintauthors{%
    \noindent\textbf{\let\@corref\@empty #2\@corref}%
    \hfill{\small\normalfont #3}\par
    {\itshape #4\par #5\par}\medskip}}
\renewcommand{\MaketitleBox}{%
  \par\begingroup
  \centering {\Large\bfseries\@title\par}\vspace{18pt}
  \begin{minipage}{\textwidth}
    \raggedright\normalsize\preprintauthors
  \end{minipage}\par\vspace{12pt}
  \ifvoid\absbox\else\unvbox\absbox\par\vspace{10pt}\fi
  \ifvoid\keybox\else\unvbox\keybox\par\vspace{10pt}\fi
  \endgroup}
\let\@elseads\@empty
\let\emailauthor\@gobbletwo
\makeatother

\newcommand{\spdHGCALcagradeight}{16.88\ensuremath{\times}}

\newcommand{\geoFaiss}{10.4\ensuremath{\times}}
\newcommand{\geoCuvs}{5.0\ensuremath{\times}}
\newcommand{\geoCagra}{4.4\ensuremath{\times}}

\newcommand{\fgcPCADeight}{7.5\,s}
\newcommand{\faissDeight}{310\,s}
\newcommand{\cuvsBFDeight}{146\,s}
\newcommand{\cagraNNDDeight}{127\,s}

\newcommand{\fgcVanillaBestPCAfactor}{26.6\ensuremath{\times}}

\newcommand{\spdSynthDthree}{55\ensuremath{\times}}
\newcommand{\spdSynthDfive}{4.8\ensuremath{\times}}

\newcommand{\hgcalRecHits}{1.16\,billion}

\newcommand{\mbdOptPCA}{3}
\newcommand{\mbdOptAxis}{2}
\newcommand{\mbdPCAdefault}{3}

\begin{document}

\begin{frontmatter}

\title{FastGraph: PCA-Subspace Binned $k$-Nearest-Neighbor Graph Construction for GPU Geometric Deep Learning}

\preprintauthor[cmu]{Aarush Agarwal\corref{cor1}}{aarusha@andrew.cmu.edu}
  {Carnegie Mellon University}{Pittsburgh, PA, USA}

\preprintauthor[kit]{Jan Kieseler}{jan.kieseler@cern.ch}
  {Karlsruhe Institute of Technology}{Karlsruhe, Germany}

\preprintauthor[cmu]{Raymond He}{rhe2@andrew.cmu.edu}
  {Carnegie Mellon University}{Pittsburgh, PA, USA}

\preprintauthor[cmu]{Matteo Cremonesi}{mcremone@andrew.cmu.edu}
  {Carnegie Mellon University}{Pittsburgh, PA, USA}

\preprintauthor[uzh]{Shah Rukh Qasim}{shah.rukh.qasim@cern.ch}
  {University of Zurich}{Zürich, Switzerland}

\cortext[cor1]{Corresponding author}

\affiliation[cmu]{organization={Carnegie Mellon University},
            city={Pittsburgh},
            state={PA},
            country={USA}}

\affiliation[kit]{organization={Karlsruhe Institute of Technology},
            city={Karlsruhe},
            country={Germany}}

\affiliation[uzh]{organization={University of Zurich},
            city={Zürich},
            country={Switzerland}}

\begin{abstract}
Dynamic graph neural networks repeatedly construct $k$-nearest-neighbor
graphs in learned latent spaces, making GPU-accelerated graph construction
an important operation in scientific point-cloud workloads.
We present \textbf{FastGraph}, a PyTorch-integrated primitive for exact
$k$NN graph construction in low-to-moderate-dimensional spaces.
FastGraph uses a PCA-subspace cell list: it fits a compact binning coordinate
system to the input tensor, uses that subspace for spatial pruning, and
evaluates all candidate distances in the original feature space. The
data-adaptive orthonormal subspace improves uniform-grid pruning in the
$d{=}4$--$10$ range evaluated around the $d{=}4$ GravNet operating point.
Point data remain GPU-resident, with small host synchronizations for control
metadata. The implementation ships with an eager-mode PCA API and an opt-in
differentiable \texttt{GravNetOp} integration. On a 5\,M-point, single-segment
detector-derived HGCAL recHit stress test,
FastGraph builds the exact graph in \fgcPCADeight\ at $d{=}8$,
$k{=}40$, achieving speedups of up to $41\times$ over FAISS-GPU,
$19\times$ over cuVS brute force, and $17\times$ over approximate
CAGRA while retaining
exact neighbor sets under the paper's distance-based convention.
\end{abstract}


\end{frontmatter}

\section{Introduction}

Graph neural networks (GNNs) have become a standard substrate for
reasoning over irregular relational data, from web-scale recommender
systems~\cite{ying2018pinsage} to particle reconstruction in
high-granularity calorimeters~\cite{Qasim2019GravNet,Kieseler2020ObjectCondensation}.
A defining architectural choice in many modern GNNs is that the graph
itself is built at training time, from a learned latent embedding, by
a $k$-nearest-neighbor query.
The dynamic kNN graph is constructed on every forward pass, and its cost can be a
substantial component of a model step. Establishing that fraction for
the target HGCAL models requires end-to-end profiling and is outside
the kernel study reported here. Importantly, the latent
dimensionality in the target application is low-to-moderate-dimensional: the original
GravNet layer for HGCAL reconstruction~\cite{Qasim2019GravNet} uses
$d{=}4$ for the learned space in which the kNN is taken. Related
multi-particle and object-condensation models used in CMS HGCAL
reconstruction~\cite{Kieseler2020ObjectCondensation,qasim2021multiparticle,bhattacharya2022gnn}
motivate evaluating the nearby $d{=}4$--$10$ range. EdgeConv~\cite{wang2019dynamic} and
ParticleNet~\cite{qu2020particlenet} also rebuild feature-space graphs,
but later layers may operate at substantially higher dimensionality and
are outside the primary regime evaluated here. This
\emph{low-to-moderate-dimensional} regime is precisely where neither generic
high-dimensional ANN libraries nor brute-force GPU $k$NN are an ideal
fit: ANN libraries introduce unfavorable overheads, and brute force
scales as $N^2 d$ regardless of how much spatial structure the latent
embedding contains.

\paragraph*{Why exact $k$NN matters in this regime}
Exact $k$NN provides a fixed distance criterion without an index-specific
recall target or quality hyperparameter. Tied distances may still admit
multiple valid exact neighbor sets unless a tie-breaking rule is fixed. Approximate
construction changes the selected edge set, and graph-learning studies
show that edge perturbations can alter downstream predictions
substantially~\cite{zugner2018adversarial,klicpera2019diffusion}.
The present work does not establish that approximate-$k$NN errors
produce a measurable HGCAL physics systematic; that question requires
an end-to-end model study. We instead evaluate the narrower engineering
case for exact construction when it is competitive in the target
low-to-moderate-dimensional regime.

The classical low-dimensional accelerator for this regime is the
\emph{cell list}~\cite{verlet1967,howard2016neighbor}: place points in a uniform
grid, expand neighboring bins around each query until the best-$k$
radius is certified, and evaluate distances only for candidates in
the visited bins. A full $d$-dimensional grid scales as
$n_{\text{bins}}^d$, so practical GPU implementations limit the binning
grid to a small $k$ even when the input has $d>k$ coordinates. FastGraph
addresses this constraint with a data-adaptive choice of binning coordinates:
it bins in the top-$k$ principal subspace while retaining full-dimensional
distances for candidate evaluation. Since the partial projection
$V_k V_k^\top \preceq I_d$ is contractive, projected-space
termination remains a valid lower bound on true-space distance, so
the algorithm preserves exactness for supported input dimensions. The PCA formulation is
evaluated at $d{=}4$--$10$; $d{=}2$--$3$ are axis-aligned controls for
which the implementation short-circuits before PCA. At $d{=}8$,
$N{=}5$\,M, $k{=}40$, FastGraph builds the kNN graph in
\fgcPCADeight\ versus \faissDeight\ for exact FAISS-GPU,
\cuvsBFDeight\ for cuVS brute force, and \cagraNNDDeight\ for
CAGRA's approximate NN-Descent build. Across $d{=}2$--$10$ at this
$(N,k)$ point, geomean speedups are \geoFaiss, \geoCuvs, and
\geoCagra\ respectively.

The library is shipped as a PyTorch extension. The axis-aligned primitive
remains callable from TorchScript, while the PCA wrapper uses eager-mode
\texttt{torch.pca\_lowrank}. Gradients flow through the continuous outputs of the kernel
(selected squared distances, query coordinates) while the hard
neighbor indices are treated as constants in the backward pass.
\texttt{GravNetOp}~\cite{Qasim2019GravNet} exposes PCA-subspace neighbor
selection through \texttt{use\_pca=True} in eager mode.

\section{Related Work}
\label{sec:related-work}

\paragraph*{Exact GPU $k$NN}
Modern GPUs make the inner-loop distance computation of exact $k$NN
embarrassingly parallel. FAISS~\cite{johnson2019billion} provides a
highly tuned \texttt{IndexFlatL2} exact brute-force GPU index together
with the fast $k$-selection primitives that downstream methods reuse.
RAPIDS \texttt{cuVS} exposes a comparable exact brute-force GPU
implementation that we use as the strongest exact baseline in this
work; KeOps~\cite{feydy2020fast} provides symbolic matrix reductions
with automatic differentiation that can express exact $k$NN without
materializing the full distance matrix. Direct exact GPU $k$NN-graph
construction has also been studied in the multi-GPU and cluster
settings~\cite{dashti2013efficient,komarov2014fast}. PyTorch Geometric
exposes the widely used \texttt{knn\_graph} interface through its
\texttt{torch\_cluster} extension~\cite{fey2019pyg}; we treat it as a
framework interface rather than a distinct algorithmic baseline.
FastGraph is not
the first exact GPU $k$NN library; it specializes the accelerator for
the small-$d$ exact graph-construction setting that arises in dynamic
GNN layers.

\paragraph*{Cell lists and spatial binning}
The cell list is the canonical low-dimensional accelerator: it dates
to Verlet's molecular-dynamics neighbor-finding scheme and underpins
modern MD libraries such as HOOMD-blue. The construction is the same
across communities---uniform grid in $d$ dimensions, sort points by
bin, expand a concentric hypercube of bins around each query until
the best-$k$ radius is certified---but the choice of $d$ has been
constrained in practice to two or three axis-aligned dimensions
because the bin tensor scales as $n_{\text{bins}}^d$.
Qasim's PhD thesis~\cite{qasim2023thesis} discusses the practical grid
dimensionality limit in the GNN context. FastGraph uses a principal subspace
for the grid coordinates; an axis-aligned first-column choice is retained only
as an internal ablation.

\paragraph*{PCA-projected indexing}
Using a principal-component projection as a coordinate system for
spatial indexing has a long history; Sproull's $k$-d tree
refinements~\cite{sproull1991refinements} already noted that
splitting on the direction of maximum variance yields more balanced
partitions than splitting on a coordinate axis. Principal-axis tree
variants use a similar idea for CPU-side neighbor search. FastGraph's
PCA-subspace binning differs in
two ways: the projection is partial (top-$k$ axes) rather than full,
and the candidate-distance evaluation that follows the pruning step is
performed in the \emph{original} $d$-dimensional space, preserving
exactness. The contraction $V_k V_k^\top \preceq I_d$ is what makes
that combination valid (Section~\ref{sec:method}).

\paragraph*{Three-dimensional spatial-acceleration $k$NN}
A separate line of work targets exact GPU $k$NN through dedicated
spatial-acceleration structures that exploit the GPU's
ray-tracing hardware or BVH machinery. RTNN~\cite{wald2019rtnn},
RT-kNNS Unbound~\cite{evangelou2021rtknn}, and
Arkade~\cite{nagarajan2023arkade} retarget the OptiX BVH unit for
neighbor search and report large speedups over shader-core baselines
in 3D. LBVH-kNN~\cite{jakob2021optimized} (\texttt{cupy-knn}) attains
similar gains via a left-balanced LBVH built without dedicated
RT hardware. CLOVER~\cite{kamel2025clover} introduces a GPU-native,
Voronoi spatio-graph approach to exact $k$NN and reports
$4\times$ over an ``optimized grid'' baseline and
$230\times$ over FAISS on low-dimensional spatial data. All of these methods are
implemented and evaluated for two- or three-dimensional inputs: Arkade
notes that RT cores build BVHs only on three-dimensional data, CLOVER's
reference implementation asserts \texttt{D == 3}, and LBVH-kNN is
evaluated exclusively in 3D. We include a direct head-to-head against
CLOVER at $d{=}3$ in Section~\ref{sec:clover}, but the released methods
do not provide a $d{=}4$--$10$ path comparable to the one evaluated here.

\paragraph*{Approximate GPU $k$NN}
A parallel body of work has produced GPU-resident approximate nearest
neighbor (ANN) systems. SONG~\cite{zhao2020song} decomposes graph
traversal into GPU-friendly stages; GGNN~\cite{ggnn} builds a
hierarchical $k$NN graph for index construction and search;
CAGRA~\cite{ootomo2023cagra}, distributed as part of RAPIDS
\texttt{cuVS}, builds a high-quality graph index with an IVF-PQ or
NN-Descent initialization. ANN-Descent variants have also been adapted
to reduce memory traffic during construction~\cite{wang2021large}.
These systems expose a tunable recall--throughput tradeoff for
high-throughput vector retrieval; FastGraph instead targets exact,
small-$d$ graph construction. We benchmark CAGRA and GGNN as approximate
comparators (Section~\ref{sec:recall}) but do not include CPU-only ANN
systems (HNSW, ScaNN, Annoy)~\cite{malkov2020hnsw,guo2020scann,bernhardsson2013annoy}
in the timing tables: the point tensors and large intermediate arrays remain
GPU-resident, and a CPU-side detour for bulk data would not be a fair comparator.

\paragraph*{Dynamic latent-graph models in geometric deep learning}
Dynamic kNN-graph layers in learned latent spaces are a recurring
ingredient in geometric deep learning. DGCNN/EdgeConv recomputes the
graph in feature space at each layer~\cite{wang2019dynamic};
ParticleNet adapts the same idea to jet
tagging~\cite{qu2020particlenet}; GravNet/GarNet
layers~\cite{Qasim2019GravNet} and the object condensation
loss~\cite{Kieseler2020ObjectCondensation} are deployed for HGCAL
reconstruction~\cite{qasim2021multiparticle,bhattacharya2022gnn}; the
HEP.TrkX/Exa.TrkX line of work uses learned-embedding graphs for
tracking~\cite{ju2021performance,lazar2022accelerating,choma2020track}.
FastGraph is best framed as an exact substrate for the kNN-graph layer
inside such models---not a new model class.

\section{The PCA-Subspace Binned $k$NN Algorithm}
\label{sec:method}

In the algorithm and complexity analysis, $K$ denotes the requested
neighborhood size (including self), and $k$ denotes the binning-subspace
dimension. Experimental figure labels use the conventional $k$ for
neighborhood size; binning dimensions are identified explicitly.

\subsection{Phase 1: Subspace estimation}
\label{sec:phase1}

Given the input coordinates
$X \in \mathbb{R}^{N \times d}$ for the current call, FastGraph samples
up to 50,000 rows, centers that sample, and computes the top-$k$ principal
axes with \texttt{torch.pca\_lowrank}. The current release fits one
projection for the complete coordinate tensor rather than a separate
projection for each \texttt{row\_splits} segment. The number of
binning axes $k$ is a user-facing hyperparameter; we default to
$k = \mbdPCAdefault$ for the HGCAL workloads in this paper, justified
empirically in the ablation of Section~\ref{sec:mbd-ablation}. The
resulting projection matrix $V_k \in \mathbb{R}^{d \times k}$
satisfies $V_k^\top V_k = I_k$ and $V_k V_k^\top \preceq I_d$. The
projected coordinates used for binning are $Y = X V_k$. PCA estimation
is performed once per call on the GPU and is included in the reported
FastGraph timing. The random subsample follows PyTorch's active random
number-generator state.

\paragraph*{Short-circuit at $d \le k$}
When the input dimensionality $d$ does not exceed the binning
dimensionality $k$, FastGraph skips PCA estimation entirely and
binds the binning axes to the raw input axes; the projection
$V_k$ reduces to the identity (up to permutation), $Y = X$, and the
algorithm behaves exactly as a classical axis-aligned cell list. With
the default $k = \mbdPCAdefault$, the $d \in \{2, 3\}$ gains therefore
come from the underlying cell-list machinery; the PCA contribution
begins at $d > k$, which covers the $d{=}4$--$10$ HGCAL latent-space
range.

\subsection{Phase 2: Binning in the projected subspace}
\label{sec:phase2}

We partition $Y$ with a uniform grid whose scalar width $w$ is shared
across the $k$ projected axes. The division count is set heuristically
to balance bin density against neighborhood reach,
\[
  n_{\text{bins}} = \left(\frac{32\,n_{\text{elems}}}{K}\right)^{\frac{1}{k}},
\]
where the released implementation uses the historical occupancy proxy
\[
  n_{\text{elems}}=\lfloor\max(\texttt{row\_splits})/
  |\texttt{row\_splits}|\rfloor+1,
\]
and $K$ is the requested neighborhood size. The integer division count
is clamped to $[5,30]$. For the one-segment
benchmark input \texttt{row\_splits}=$[0,N]$, this proxy is approximately
$N/2$; it is a performance heuristic and does not enter the exactness
condition. Points are sorted by bin index, and cumulative offsets are
stored so that each bin can be addressed as a contiguous slice of the
sorted point array. Row splits, which delineate the per-event
boundaries inside a batched tensor, are honored by the binning kernel
so that no query bin spans events.

\subsection{Phase 3: Per-query search with full-dim distances}
\label{sec:phase3}

For each query $x_q$, the search starts in the bin containing
$y_q = x_q V_k$ and expands through successive shells of $k$-D
hypercube bins. For every candidate $u$ in a visited bin the
\emph{full-dimensional} squared Euclidean distance
$\|x_q - x_u\|^2 = \sum_{i=1}^{d}(x_{q,i} - x_{u,i})^2$ is computed
(not the $k$-dimensional projected distance). The current best
neighbor radius $r_K$ and a fixed-size top-$K$ candidate buffer are
maintained incrementally.

\subsection{Why this is exact: the contraction argument}
\label{sec:exactness}

For finite inputs with valid row splits, a supported projection dimension
$2 \le k \le d$, and at least $K$ eligible points in each segment (or in
its direction-masked subset), consider the standard cube-radius termination
test applied to bins in the projected subspace while all candidate distances
are evaluated in the full input space. The
exactness of this test rests on a single observation about the partial
projection. For any two points $x_a, x_b \in \mathbb{R}^d$,
\[
  \|y_a - y_b\|^2
  = (x_a - x_b)^\top V_k V_k^\top (x_a - x_b)
  \le \|x_a - x_b\|^2,
\]
because $V_k V_k^\top$ has eigenvalues in $\{0,1\}$ and is therefore
$\preceq I_d$. Equivalently, the projection is contractive: the
projected distance never overestimates the true distance.

After visiting all bins within the current hypercube shell of radius
$s$, any unvisited bin lies at projected distance at least
$w \cdot s$ from $y_q$, where $w$ is the scalar width shared by all
projected axes in the released implementation. Thus any unvisited point
satisfies $\|y_q - y_u\|^2 \ge (w \cdot s)^2$. Combined with the
contraction above, $\|x_q - x_u\|^2 \ge (w \cdot s)^2$, so
once the top-$K$ buffer has filled and $(w \cdot s)^2 > r_K^2$, no unseen
point can be closer in the true distance than the current $K$-th
neighbor. The algorithm may safely terminate; the returned $K$
neighbors are exact in exact arithmetic. This argument assumes an
orthonormal projection and does not bound floating-point errors in the
projection or pruning test. The FP32 implementation is validated separately
against brute force at tractable scale, with the distance criteria and
numerical tolerances reported in Section~\ref{sec:recall}.

\subsection{Binstepper}

The \texttt{binstepper} is the small bookkeeping utility that
enumerates the surface of the $s$-th hypercube shell around the
query bin. It linearises each shell into a flat sequence, converts
each step back into per-dimension offsets, discards interior cells,
and yields the next surface bin's flat index. The stepper carries no
algorithmic role; the kNN guarantee lives in the outer kernel of
Algorithm~\ref{alg:knn}. Its job is purely to enumerate the projected
shells in a deterministic flat order so the distance kernel can stay
focused on the full-dimensional checks. Keeping the shell enumerator
separate from the distance kernel also makes the projected walk
event-local: the same row-split offsets and contiguous bin slices can
be reused across all queries in the batch.

\subsection{Pseudocode}

Algorithm~\ref{alg:knn} makes the projected-vs-original-space
distinction explicit: the bin indices (\texttt{binIdx}) and the
stepper operate on $y$-space; the distance function
\texttt{calcDist} operates on the original $x$-space coordinates;
the termination test references the projected-space cube radius;
the contraction argument certifies the result. This separation keeps
the implementation and the proof aligned: projected-space search only
decides which bins to visit, while exactness is enforced by the
original-space distance computation.

\begin{algorithm}[!htbp]
\small
\DontPrintSemicolon
\KwIn{\texttt{coords}[$n_v$][$n_c$] (original-space $X$), $V_k$, \texttt{binIdx}[$n_v$] (projected-space bins), \texttt{binOffsets}[$n_v$][$n_b{+}1$], \texttt{binBounds}[$n_B$], \texttt{binCounts}[$n_b$], scalar \texttt{binWidth}, $n_v,K,n_c,n_b,n_B$}
\KwOut{\texttt{neighbors}[$n_v$][$K$] indices including slot 0 self by convention, $r_K^2$ values}
\ForEach{$v \in [0,n_v)$ \textbf{in parallel}}{
  \texttt{neighbors}[v][0]$\leftarrow(v,0)$; \texttt{filled}$\leftarrow1$; $(\texttt{maxIdx},\texttt{maxD2})\leftarrow(0,0)$\;
  $\texttt{totalBins}\leftarrow\prod_{d=0}^{n_b-1}\texttt{binCounts}[d]$, $\texttt{gOff}\leftarrow \texttt{totalBins}\cdot\lfloor \texttt{binIdx}[v]/\texttt{totalBins}\rfloor$\;
  \texttt{step}$\leftarrow$\texttt{BinStepper}(\texttt{binCounts}, \texttt{binOffsets}[v]$[1..n_b]$) \tcp*{operates in projected $y$-space}
  $w\leftarrow \texttt{binWidth}$\;
  \texttt{cont}$\leftarrow$ true; $s\leftarrow0$ \tcp*{$s$: current cube-shell radius (projected space)}
  \While{\texttt{cont}}{
    \texttt{step.setDistance}($s$); \texttt{cont}$\leftarrow$ false\;
    \While{$(\mathtt{off} \leftarrow \mathtt{step.step()}) \neq -1$}{
      $\texttt{idx}\leftarrow \texttt{off}+\texttt{gOff}$\;
      \If{$\texttt{idx}<0$ \textbf{or} $\texttt{idx}\ge n_B-1$}{\Continue}
      $(s_b,e_b)\leftarrow(\texttt{binBounds}[\texttt{idx}],\texttt{binBounds}[\texttt{idx}{+}1])$\;
      \If{$s_b=e_b$}{\texttt{cont}$\leftarrow$ true; \Continue}
      \For{$u\in[s_b,e_b)$}{
        \If{$u{=}v$}{\Continue}
        $\texttt{d2}\leftarrow \texttt{calcDist}(v,u)$ \tcp*{full $d$-dim $\sum_i(x_{v,i}-x_{u,i})^2$, NOT projected}
        \If{$\texttt{filled}<K$}{\texttt{neighbors}[v][\texttt{filled}]$\leftarrow(u,\texttt{d2})$;\ \If{$\texttt{d2}>\texttt{maxD2}$}{$(\texttt{maxIdx},\texttt{maxD2})\leftarrow(\texttt{filled},\texttt{d2})$}; \texttt{filled}++}
        \ElseIf{$\texttt{d2}<\texttt{maxD2}$}{\texttt{neighbors}[v][\texttt{maxIdx}]$\leftarrow(u,\texttt{d2})$;\ $(\texttt{maxIdx},\texttt{maxD2})\leftarrow\texttt{findMaxDist}(\texttt{neighbors}[v])$}
      }
      \texttt{cont}$\leftarrow$ true
    }
    \If{$\texttt{filled}{=}K$ \textbf{and} $(w\cdot s)^2>\texttt{maxD2}$}{\Break}
    $s$++
  }
}
\caption{\textbf{PCA-Subspace Binned-Select-kNN}: bins live in projected $y$-space; distances are computed in original $x$-space. Slot 0 is the query point itself, matching the FastGraph tensor convention; downstream message-passing code may drop or ignore this self edge.}
\label{alg:knn}
\end{algorithm}

\vspace{-0.6em}
\subsection{Complexity}

For a single segment, let $B = n_{\text{bins}}^{k}$ be the size of the
bin grid; for multiple segments, let $B$ denote their total allocated bin
count. Both PCA-subspace binning and the axis-aligned $k$-coordinate
control avoid a full $n_{\text{bins}}^{d}$ grid. PCA changes the choice of
binning axes rather than the grid's dimensional dependence.

\noindent\textit{Setup.} Let $S=\min(N,50{,}000)$ be the PCA sample
size and $q=\min(k+2,d)$ the low-rank solve size. For the fixed number
of randomized iterations used by \texttt{torch.pca\_lowrank}, the
subspace fit costs $O(Sdq+(S+d)q^2)$, followed by $O(Ndk)$ to project
all points. This is a sampled low-rank computation, not a full
$O(Nd^2+d^3)$ eigendecomposition. Binning, sorting, and offset
construction take $O(N \log N + B)$.

\noindent\textit{Memory.} Auxiliary memory is $O(N(d + k + K) + B)$:
sorted coordinates, projected coordinates, permutation, neighbor
indices, and bin offsets.

\noindent\textit{Per-query search.} Let $C_v$ be the number of candidate
points evaluated for query $v$, and $E_v$ the number of cell-offset tests
performed by the shell enumerator, including rejected offsets. Each
candidate requires $O(d)$ distance work and at most $O(K)$ for the
top-$K$ update; cell-offset bookkeeping costs $O(k)$ per test. The query
cost is therefore $O(kE_v+C_v(d+K))$. Uniform projected density alone
does not imply $C_v=O(K)$: many points may be close in projection but far
apart in the discarded dimensions. When pruning and bin occupancy keep
both $C_v$ and $E_v$ proportional to $K$, this reduces to $O(K(d+K))$
for fixed $k$. In the worst case, every point in a segment is evaluated,
giving $O(N_v(d+K))$ candidate work for segment size $N_v$, in addition
to shell enumeration.

\noindent\textit{Total work.} Summing across queries gives
\[
 O\!\left(Sdq+(S+d)q^2+Ndk+N\log N+B+
 \sum_{v=1}^{N}\bigl[kE_v+C_v(d+K)\bigr]\right).
\]
Queries execute in parallel on the GPU; elapsed time also depends on
load imbalance, memory traffic, and resource occupancy.

\subsection{Gradient flow and PyTorch integration}

FastGraph exposes gradients through the continuous outputs of the
kernel---the selected squared distances and the input coordinates---
while treating the discrete neighbor indices and the projection
matrix $V_k$ as constants in the backward pass. Concretely, the
forward returns $(\text{idx}, \text{dist}^2)$ and the backward
applies the chain rule with respect to $\text{dist}^2$ and the input
coordinates, yielding a subgradient of the piecewise-smooth loss
induced by the hard neighbor selection. This is the same gradient
model used by any fixed-index distance layer
(e.g.,\ \texttt{torch.cdist} followed by \texttt{topk}) and integrates
naturally into existing GNN autograd graphs. Treating $V_k$ as fixed per forward--backward pair keeps graph selection
consistent within that pair; the projection is refreshed at the next forward.
The CUDA backward uses atomic accumulation and is therefore not guaranteed
to be bitwise deterministic.

The PCA entry point is an eager-mode Python wrapper around the scripted
axis-aligned primitive and the custom CUDA operation. A minimal usage
example is:

\begin{lstlisting}[language=Python, float=htbp, caption={Minimal eager-mode PCA-FastGraph usage.}]
from fastgraphcompute import binned_select_knn_pca
from fastgraphcompute.gnn_ops import GravNetOp

# Direct call (eager mode)
idx, dist2 = binned_select_knn_pca(K=40, coords=x, row_splits=rs,
                                   max_bin_dims=3)
# idx:   (N, K) int64 neighbor indices (constant in backward)
# dist2: (N, K) float32 squared distances (differentiable)
loss = model(dist2).sum()
loss.backward()  # gradients flow through dist2 to x

# Opt-in PCA neighbor construction inside GravNetOp (eager mode)
layer = GravNetOp(in_channels=16, out_channels=32,
                  space_dimensions=8, propagate_dimensions=32,
                  k=40, use_pca=True, max_bin_dims=3)
out, idx, dist2, space = layer(features, rs)
\end{lstlisting}

\FloatBarrier
\section{Experimental Setup}
\label{sec:expsetup}

All GPU experiments use a single NVIDIA A100 80\,GB PCIe with CUDA
12.1 and PyTorch 2.5. Compared backends are exact FAISS-GPU
\texttt{IndexFlatL2}~\cite{johnson2019billion}, cuVS brute force
(also exact), CAGRA via cuVS~\cite{ootomo2023cagra}
(approximate), and GGNN~\cite{ggnn} (approximate); FastGraph is
compiled from the manuscript's \texttt{paper-release} branch.

\paragraph*{Timing protocol}
All comparative HGCAL timing plots and the headline table use the completed
matched-input, serialized \texttt{--gres=mps:100} campaign only; older
\texttt{mps:50} and exclusive-allocation rows are excluded. Every backend
has seven repetitions per completed plotted cell for FastGraph/PCA, FAISS,
cuVS BF, and GGNN; CAGRA-NN-Descent has three. Inputs are copied to a
Torch CUDA tensor before timing: FastGraph and GGNN consume that tensor
directly, FAISS uses its Torch bridge, and cuVS BF and CAGRA use a
pointer-identical CuPy DLPack view created before the timer. The timed
region therefore includes index construction and search, but excludes data
loading and host-to-device transfer for every backend. All timed regions are
bracketed by device synchronization. Figures and scalar comparisons report
medians across successful repetitions; error bars show the interquartile
range.

\paragraph*{CAGRA build path}
The default CAGRA IVF-PQ build path fails in the 5\,M-point timing campaign at $d \ge 5$
with a RAFT assertion about invalid or duplicated neighbor nodes. We did
not isolate the internal cause of this failure. We therefore use CAGRA's
\texttt{build\_algo="nn\_descent"}
path for the headline timing campaign; it avoids the PQ initialization,
succeeds across the full $d{=}2$--$10$ sweep, and is the robust full-sweep
CAGRA timing baseline. The archived recall diagnostic predates that campaign
and is labeled separately as default-build CAGRA.

\paragraph*{The \texttt{max\_bin\_dims} hyperparameter}
The binning subspace size $k$ is exposed as
\texttt{max\_bin\_dims}. The benchmark binary and evaluated kernel surface
provide compile-time template instantiations for $k \in \{2,3,4,5\}$; we
use $k=\mbdPCAdefault$ as the default for HGCAL and ablate that surface in
Section~\ref{sec:mbd-ablation}. The immutable source release also contains
experimental $k=6,7$ dispatches used to probe the design boundary, but they
are not part of the recommended headline configuration; $k=7$ exceeds the
current per-block register budget for some configurations. The empirical
optimum on HGCAL data is well inside the evaluated range.

\paragraph*{Benchmark data}
The detector-derived benchmark matrix comprises \hgcalRecHits\ CMS HGCAL
simulation recHits concatenated across events, with ten stored feature
columns per hit. The harness deterministically takes the first $N$ rows
and first $d$ columns; the available artifact does not identify the column
names, units, preprocessing, or normalization. Increasing $d$ therefore
changes both the number and identity of included features, and PCA remains
sensitive to their numerical scales. Each timed slice is passed to the
implementations as one segment rather than preserving event boundaries. This construction
is a large-$N$ stress test of the kernels on detector-derived features; it
is not an event-batched end-to-end GNN workload or a learned-latent sample.
Synthetic comparisons use isotropic Gaussian
$\mathcal{N}(0, I_d)$ samples; this matches the benchmark protocol
used in this paper and isolates the effect of dimensionality from the
feature-scale effects present in the detector-derived matrix.

\section{Performance Comparisons}

\subsection{Headline benchmark: Dimensional scaling on detector data}
\label{sec:headline}

\begin{figure}[!htbp]
\centering
  \begin{minipage}[c]{0.56\textwidth}
  \centering
  \includegraphics[width=\linewidth]{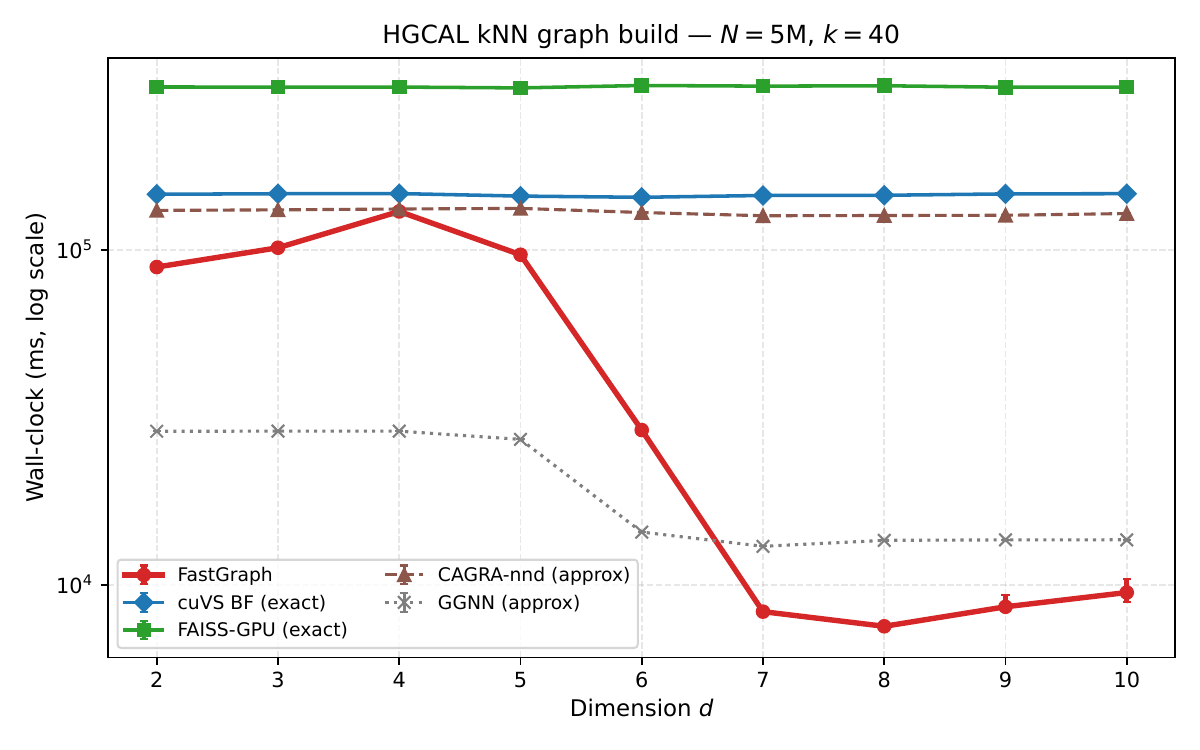}
  \end{minipage}
  \hfill
  \begin{minipage}[c]{0.42\textwidth}
  \centering
  \scriptsize
  \setlength{\tabcolsep}{1.8pt}
  \begin{tabular}{ccccc}
  \toprule
  $d$ & FAISS & cuVS & CAGRA & GGNN \\
  \midrule
  2  & 3.45$\times$  & 1.65$\times$  & 1.48$\times$  & 0.32$\times$ \\
  3  & 3.02$\times$  & 1.45$\times$  & 1.30$\times$  & 0.28$\times$ \\
  4  & 2.35$\times$  & 1.13$\times$  & 1.02$\times$  & 0.22$\times$ \\
  5  & 3.15$\times$  & 1.50$\times$  & 1.38$\times$  & 0.28$\times$ \\
  6  & 10.71$\times$ & 4.96$\times$  & 4.47$\times$  & 0.50$\times$ \\
  7  & 37.17$\times$ & 17.52$\times$ & 15.24$\times$ & 1.57$\times$ \\
  8  & 41.23$\times$ & 19.41$\times$ & 16.88$\times$ & 1.81$\times$ \\
  9  & 35.68$\times$ & 17.13$\times$ & 14.79$\times$ & 1.59$\times$ \\
  10 & 32.36$\times$ & 15.55$\times$ & 13.56$\times$ & 1.44$\times$ \\
  \midrule
  geo. & 10.4$\times$ & 5.0$\times$ & 4.4$\times$ & 0.64$\times$ \\
  \bottomrule
  \end{tabular}
  \end{minipage}
  \caption{Headline benchmark on detector data at $N{=}5$\,M, $k{=}40$.
  Left: dimensional scaling of wall-clock time. FastGraph is the
  fastest among the tested dimension-general exact GPU baselines across
  the full $d{=}2$--$10$ range.
  Right: speedups of FastGraph over each backend at the same operating
  point. FAISS and cuVS BF are exact; CAGRA-nnd and GGNN are approximate
  baselines. Recall was not measured for these timed configurations;
  Section~\ref{sec:recall} reports a separate diagnostic experiment.}
  \label{fig:headline_dim}
\end{figure}

The right-hand table in Figure~\ref{fig:headline_dim} reports the cross-method speedups
of FastGraph at the headline operating point. FastGraph is the fastest
among the tested dimension-general \emph{exact} GPU $k$NN baselines at every dimension we measure
and has a \geoCagra\ geomean advantage over the approximate
CAGRA-NN-Descent build. The GGNN approximate baseline does win on raw
wall-clock at low $d$. Section~\ref{sec:recall} reports recall separately
for the available diagnostic configurations; these do not establish a
matched-quality comparison for the headline timings. The relative
advantage is widest at $d{=}7$--$10$. Geomean across $d{=}2$--$10$ at this
operating point is \geoFaiss\ over FAISS, \geoCuvs\ over cuVS BF
(the strongest exact GPU baseline), and \geoCagra\ over
CAGRA-NN-Descent. Section~\ref{sec:mbd-ablation} separately reports
the internal axis-aligned ablation that motivated the PCA-subspace
design; the headline table therefore reports only the PCA-subspace
method.

FastGraph's runtime drops markedly from $d{=}5$ through $d{=}7$,
while FAISS and cuVS BF remain approximately flat across the measured
dimensional range. The wider speedup therefore comes primarily from
faster FastGraph execution, rather than increasing brute-force timings.
Because each successive $d$ appends a specific stored feature column,
this change cannot be attributed to dimensionality or PCA anisotropy
alone without feature-subset and normalization controls.

\subsection{Dataset-size scaling}

\begin{figure}[!htbp]
\centering
  \begin{minipage}{0.49\textwidth}
  \centering
  \includegraphics[width=\linewidth]{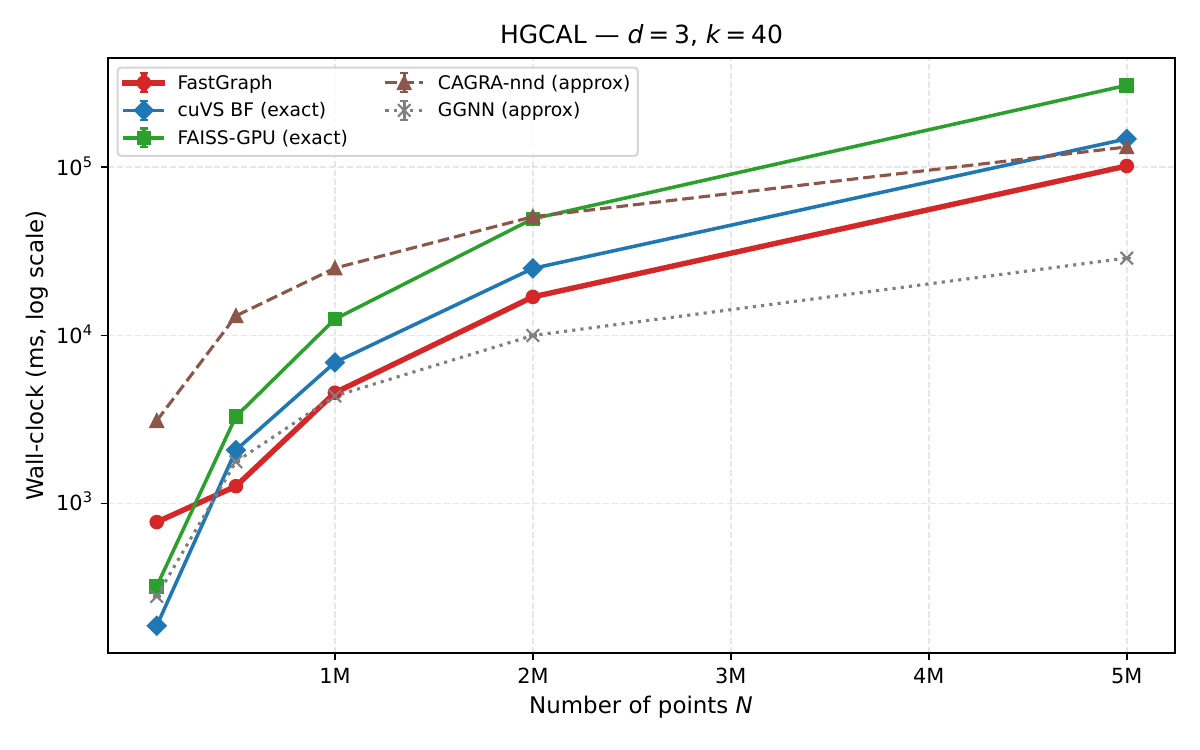}
  \end{minipage}
  \hfill
  \begin{minipage}{0.49\textwidth}
  \centering
  \includegraphics[width=\linewidth]{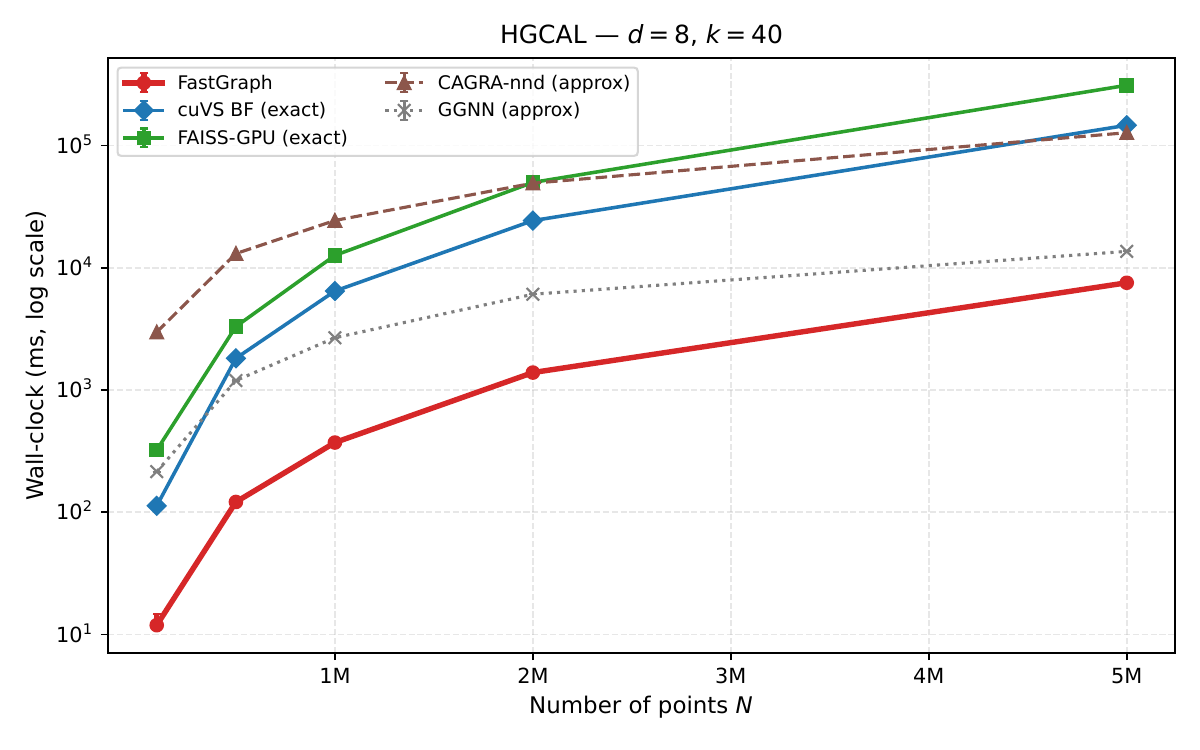}
  \end{minipage}
  \caption{Dataset-size scaling on HGCAL at $k{=}40$ for $d{=}3$
  (left) and $d{=}8$ (right). FastGraph maintains a speed advantage
  over exact baselines most clearly in the higher-dimensional
  PCA-favorable regime.}
  \label{fig:size_scaling}
\end{figure}

Figure~\ref{fig:size_scaling} reports dataset-size scaling at fixed
$k{=}40$ for $d{=}3$ and $d{=}8$ on the common five-point
\texttt{mps:100} grid. Separation is strongest in the PCA-favorable
regime.

\begin{figure}[!htbp]
\centering
  \includegraphics[width=0.95\textwidth]{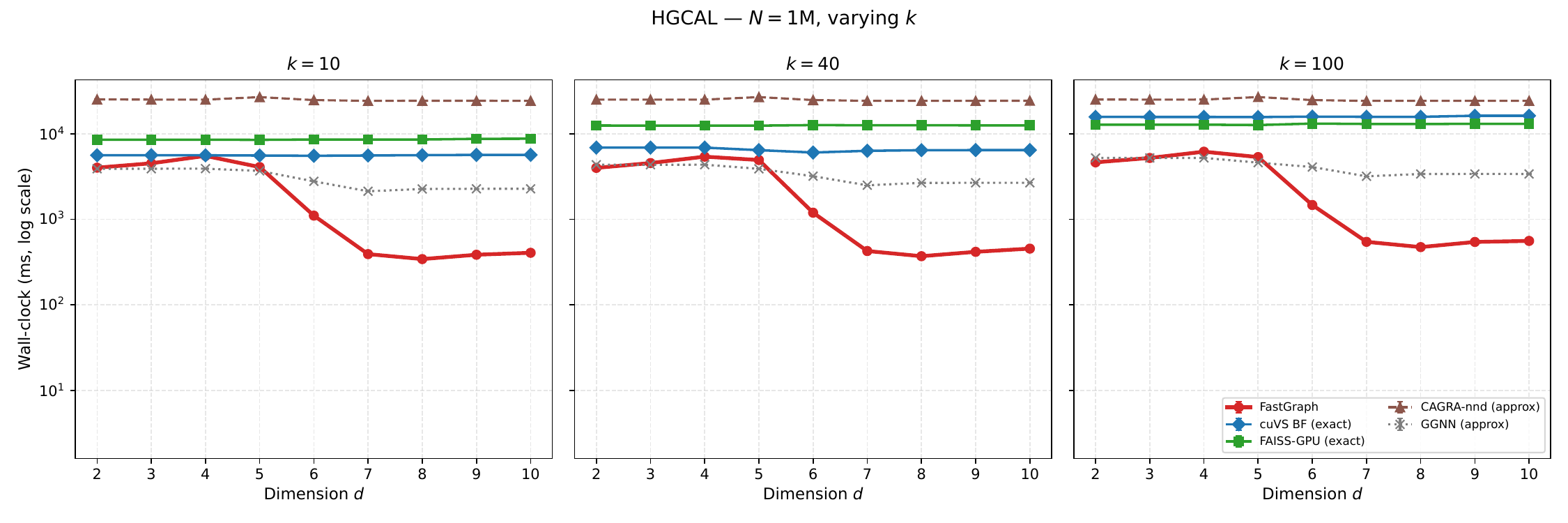}
  \caption{Wall-clock at $N{=}1$\,M for $k\in\{10, 40, 100\}$ across
  the full dimensional range. The qualitative pattern of
  Figure~\ref{fig:headline_dim} is preserved across $k$.}
  \label{fig:k_comparison}
\end{figure}

Figure~\ref{fig:k_comparison} checks that the headline dimensional
trend is not an artifact of the single $k{=}40$ operating point.
Across $k\in\{10,40,100\}$, the low-dimensional constant-factor
regime and the larger $d{\geq}6$ speedups remain visible.

\subsection{Controlled synthetic benchmarks}
\label{sec:synthetic}

\begin{figure}[!htbp]
\centering
  \includegraphics[width=0.95\textwidth]{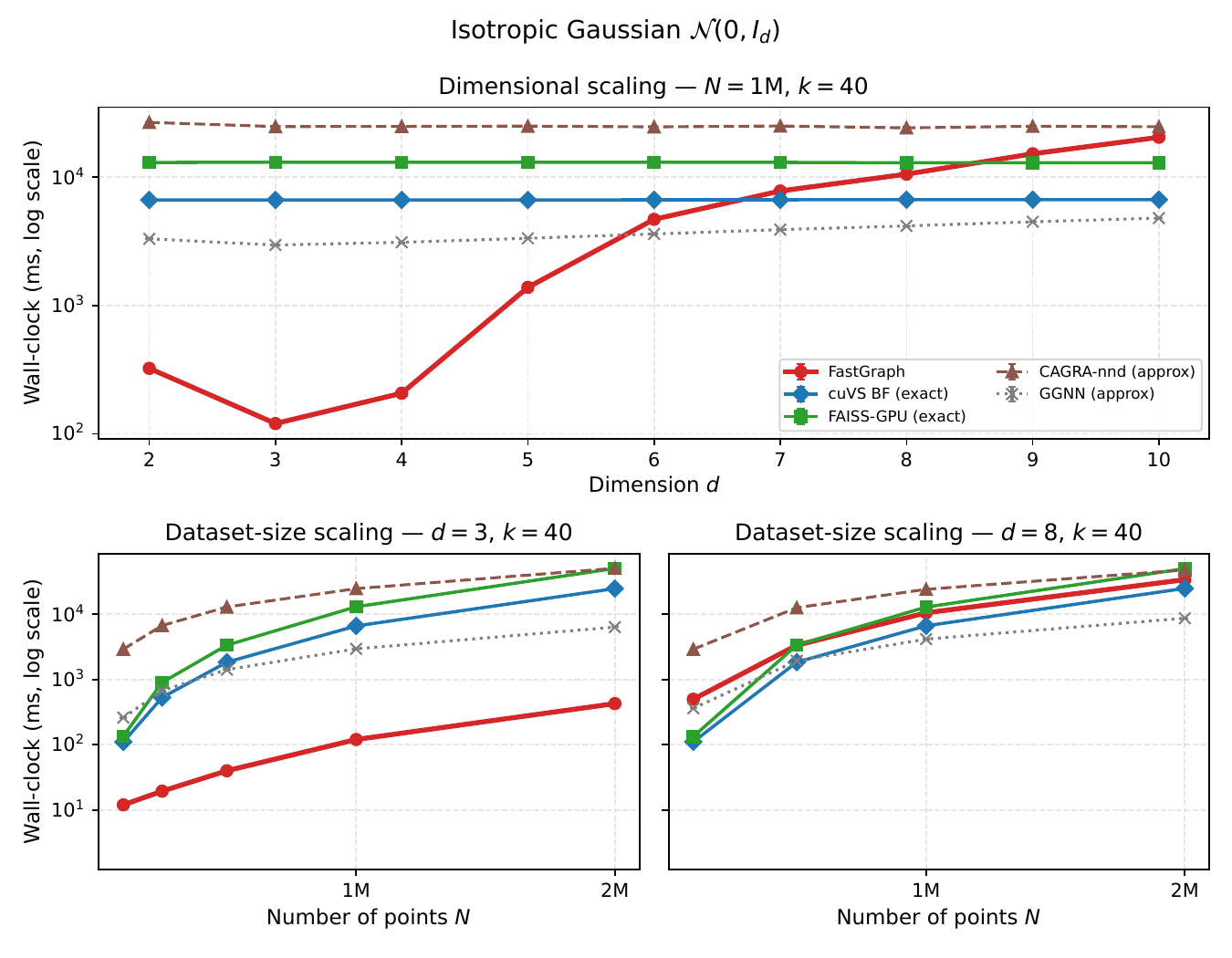}
  \caption{Controlled synthetic Gaussian benchmarks. Top:
  dimensional scaling at $N{=}1$\,M, $k{=}40$. Bottom:
  dataset-size scaling at $d{=}3$ and $d{=}8$, $k{=}40$.}
  \label{fig:synth_summary}
\end{figure}

On isotropic Gaussian $\mathcal{N}(0, I_d)$ data, FastGraph wins
decisively at low dimensions, while dense brute-force GPU matmul
(cuVS BF) becomes competitive as $d$ grows. This is consistent with the
algorithmic interpretation: the PCA step has no anisotropy to exploit
on isotropic data, so the gain comes from ordinary cell-list pruning
rather than from a favorable learned basis. At low $d$, cell-list
pruning still dominates the dense $\mathcal{O}(N^2 d)$ matmul of
the exact GPU baselines, giving FastGraph a $\spdSynthDthree$
speedup over cuVS BF at $d{=}3$ and a $\spdSynthDfive$ speedup at
$d{=}5$ (Table~\ref{tab:synth-dim}). As $d$ grows the number of
neighboring cells grows combinatorially, the binning savings
collapse, and dense matmul on a well-tuned BLAS path is close to
optimal on isotropic data. Figure~\ref{fig:synth_summary} shows the
full dimensional sweep and two dataset-size scans at $d{=}3$ and $d{=}8$.

\begin{table}[!htbp]
\centering
\caption{FastGraph speedup over GPU baselines on synthetic Gaussian
$\mathcal{N}(0, I_d)$ data at $N{=}10^6$, $k{=}40$ (median of
3 reps on the fixed-seed synthetic Gaussian benchmark set used
throughout this paper). The HGCAL detector-data wins at $d \geq 6$
(Figure~\ref{fig:headline_dim}) are not reproduced on isotropic
data, showing that the detector-data result does not follow from
dimensionality and cell-list binning alone.}
\label{tab:synth-dim}
\begin{tabular}{lrrr}
\toprule
$d$ & vs cuVS BF & vs FAISS-GPU & vs CAGRA-nnd \\
\midrule
3 & 55.2$\times$ & 108.6$\times$ & 205.8$\times$ \\
5 &  4.8$\times$ &   9.4$\times$ &  18.0$\times$ \\
7 &  0.8$\times$ &   1.7$\times$ &   3.2$\times$ \\
\bottomrule
\end{tabular}
\end{table}

The implication for the headline claim is narrower: the HGCAL detector
wins at $d{=}6$--$10$ (Section~\ref{sec:headline}) are consistent with
PCA exploiting structure absent from isotropic synthetic data, but the
current feature-prefix experiment does not isolate dimensionality,
feature identity, units, and anisotropy from one another.

\subsection{Memory footprint}
\label{sec:memory}

\begin{figure}[!htbp]
\centering
  \includegraphics[width=0.95\textwidth]{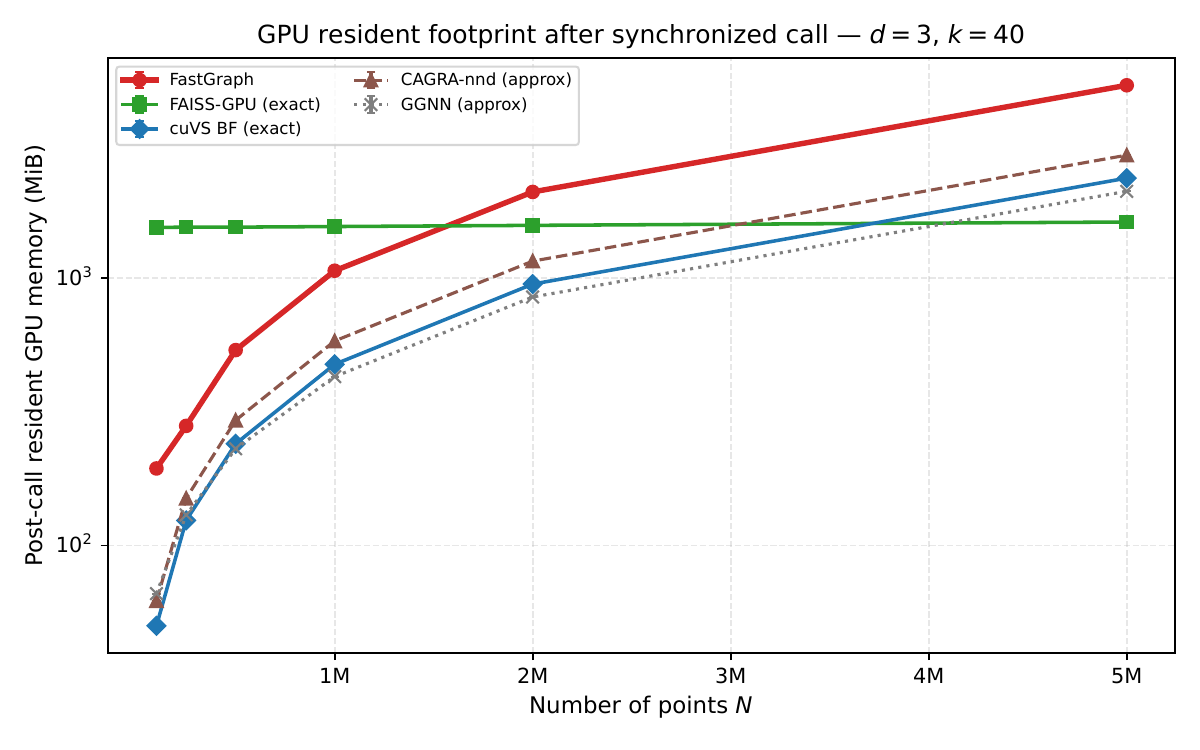}
  \caption{Post-call GPU memory at $d{=}3$, $k{=}40$. Each value is
  the change in device-wide NVML used memory between a pre-call baseline
  and a synchronized post-call sample while the caller still holds the
  backend's outputs. Values include live outputs plus allocator and index
  state; they are not transient peak-workspace measurements.}
  \label{fig:memory_footprint}
\end{figure}

\paragraph*{Resident memory}
At $N{=}5$\,M, $d{=}3$, $k{=}40$, the synchronized post-call delta is
about $5.15$\,GiB for FastGraph versus $1.58$--$2.82$\,GiB for the other
backends. FastGraph returns int64 indices and float32 distances, whose
$N{\times}k$ tensors alone occupy about $2.24$\,GiB ($2.40$\,GB) at this point; the
remainder includes live tensors and allocator state retained after the
call. CAGRA and GGNN return different output representations, so this is
an as-used resident-footprint comparison rather than an output-normalized
workspace comparison.

\paragraph*{Approximate-backend interpretation}
The plot reports each backend at the configuration used elsewhere in
this paper. FAISS-GPU \texttt{IndexFlatL2} is exact; cuVS BF is
exact. CAGRA-nn-descent and GGNN are approximate, and recall is not
measured for these timed configurations. We report the memory each
backend uses \emph{as benchmarked}; the output and index representations
differ, so this is not an output- or quality-normalized workspace comparison.

\paragraph*{Measurement methodology}
Memory is reported from synchronized before/after NVML samples. A
Python-thread poller was also recorded, but several external backends hold
the Python GIL and starve that poller during the call; those transient
samples are therefore excluded rather than presented as comparable peak
measurements.

\subsection{Binning dimensionality ablation}
\label{sec:mbd-ablation}

The binning subspace size $k$ (\texttt{max\_bin\_dims}) is the most
important FastGraph hyperparameter. We set the default to
$k{=}\mbdPCAdefault$ because it is the fastest PCA-subspace setting at
the reference cell $d{=}8$, $N{=}5$\,M, $k_{\text{nn}}{=}40$.
Table~\ref{tab:mbd-ablation} compares this choice with the
axis-aligned cell list over the evaluated kernel surface
$k \in \{2,3,4,5\}$, using medians over three repetitions.

\begin{table}[!htbp]
\centering
\caption{Effect of the binning subspace size $k$ on FastGraph
wall-clock at $d{=}8$, $N{=}5$\,M, $k_{\text{nn}}{=}40$, median of
three repetitions. The axis-aligned variant's optimum within
the evaluated kernel surface is $k{=}\mbdOptAxis$; the PCA-subspace
variant's runtime optimum, and therefore the default used in the
paper, is $k{=}\mbdOptPCA$.}
\label{tab:mbd-ablation}
\small
\begin{tabular}{cccc}
\toprule
$k$ (bin dims) & axis (s) & PCA (s) & Speedup (axis/PCA time) \\
\midrule
2 & 199.37 & 11.39 & 17.5$\times$ \\
3 & 204.65 &  \textbf{7.48} & \textbf{27.4}$\times$ \\
4 & 353.78 & 11.58 & 30.6$\times$ \\
5 & 242.90 & 28.10 &  8.6$\times$ \\
\bottomrule
\end{tabular}
\end{table}

Both variants are non-monotonic in $k$, but their optima differ: the
axis-aligned cell list is fastest at $k{=}\mbdOptAxis$, while the
PCA-subspace variant is fastest at $k{=}\mbdOptPCA$. The largest
matched-$k$ speedup occurs at $k{=}4$, but $k{=}3$ is the setting used
by FastGraph because it minimizes the PCA-subspace wall-clock. Across
the sweep, PCA-subspace binning is $8.6$--$30.6\times$ faster at
matched $k$, and comparing each variant at its own optimum gives a
\fgcVanillaBestPCAfactor\ speedup.

We expose $k$ as a user-facing hyperparameter rather than hard-coding
it: the optimum is data-dependent (a dataset with intrinsic
dimensionality near 2--3 prefers small $k$; one with variance spread
more uniformly prefers $k{=}4$ or $5$), and the same kernel binary
supports all four instantiations. The default
$k{=}\mbdPCAdefault$ is HGCAL-tuned and should be revalidated when
deploying FastGraph on datasets with different intrinsic dimension.

\subsection{Three-dimensional baseline: CLOVER head-to-head}
\label{sec:clover}

CLOVER~\cite{kamel2025clover} provides a direct comparison with a
3D-only exact-GPU $k$NN method. Its released implementation requires
$d=3$; we compare it with FastGraph on the first three stored HGCAL
feature columns and on two synthetic distributions. The HGCAL column
semantics are unavailable, so these inputs are treated as numerical
feature triples rather than identified physical spatial coordinates.

\begin{figure}[!htbp]
\centering
  \includegraphics[width=0.9\textwidth]{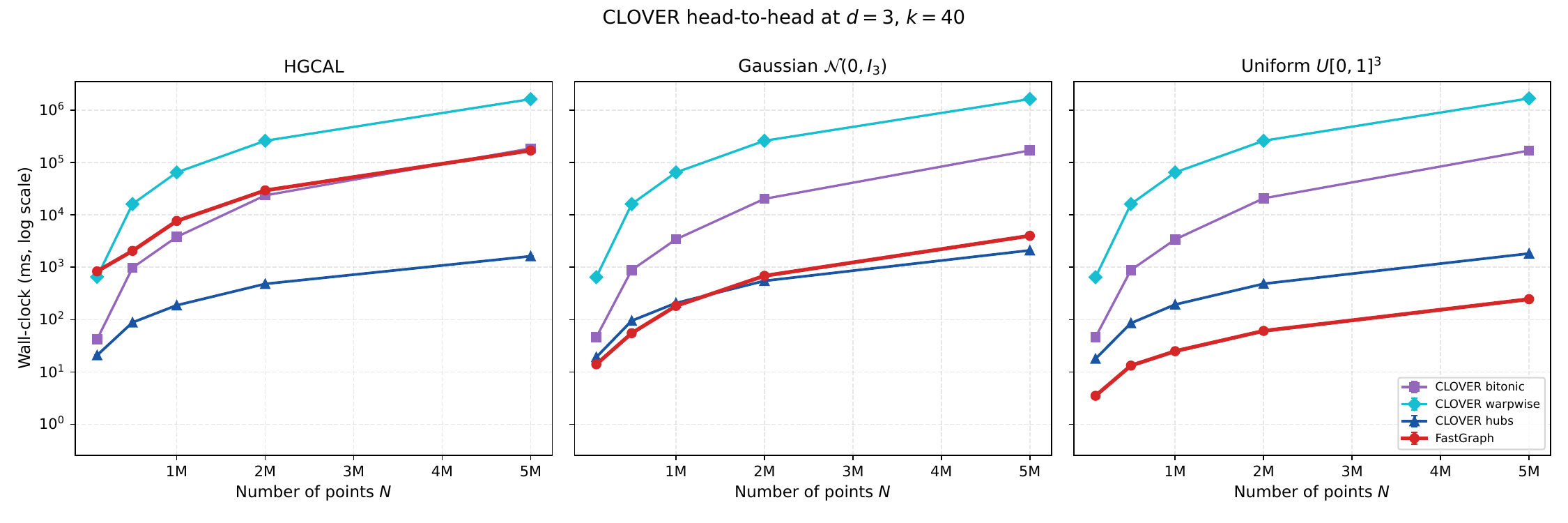}
  \caption{FastGraph vs CLOVER at $d{=}3$, $k{=}40$, across three
  data distributions: HGCAL recHits (left), synthetic Gaussian
  $\mathcal{N}(0, I_3)$ (center), synthetic uniform on $[0,1]^3$
  (right). CLOVER is only implemented for $D{=}3$, so $d{=}3$ is the
  only dimensionality at which a direct comparison is possible.}
  \label{fig:clover_d3}
\end{figure}

The result splits along the structure of the input distribution.
On uniform-cube data --- the regime where a uniform-grid cell list is
near-optimal --- FastGraph is 5--8$\times$ faster than CLOVER hubs.
On isotropic Gaussian data the two methods are within roughly a
factor of two. On anisotropic HGCAL recHits, CLOVER's hubs index is
much faster at $d{=}3$: with the default $k_{\rm bin}{=}3$,
FastGraph short-circuits to the ordinary axis-aligned cell-list path,
while CLOVER uses a 3D density-adaptive spatio-graph. This is a
favorable regime for CLOVER and an intentionally out-of-scope regime
for FastGraph's PCA-subspace contribution.

The deployment regime targeted by FastGraph is the \emph{learned latent
space} of a dynamic-graph network.
The original GravNet formulation uses a four-dimensional learned space
for neighbor construction~\cite{Qasim2019GravNet}; this work evaluates
that point and nearby low-to-moderate dimensions through $d{=}10$.
CLOVER, LBVH-kNN, and the RT-core methods considered here do not provide
a released $d \geq 4$ implementation for a direct comparison.

\subsection{Recall evaluation}
\label{sec:recall}

We use an archived broad recall sweep against a brute-force reference on
HGCAL data up to $N{=}5\times 10^5$. This diagnostic invokes the
axis-aligned \texttt{binned\_select\_knn} control rather than the PCA wrapper.
Its CAGRA rows use the default graph build and \texttt{itopk\_size}=512;
its GGNN rows use $\tau_{\text{query}}=0.8$. This archived experiment is
separate from the headline timing campaign, which uses CAGRA--NN-Descent.
We use these archived outputs only to report recall; none of their timing
values enters the performance comparisons.

Because multiple recHits can have tied distances in the selected feature
space, neighbor indices are not unique.
We use \emph{distance-based recall}, counting a predicted neighbor as correct
when its squared distance $\widehat{D}_{v,j}$ satisfies
\[
 \widehat{D}_{v,j} \le (1+10^{-4})D_{v,(K)}+10^{-10},
\]
where $D_{v,(K)}$ is the $K$-th brute-force squared distance, including
self. Recall is the fraction of the $NK$ returned entries satisfying
this threshold. FastGraph and its reference use element-wise squared
$\ell_2$ distances, which avoid the cancellation that can affect coincident-hit
cases under the BLAS expansion. FAISS-GPU \texttt{IndexFlatL2} is exact under
its own L2 kernel and is shown as such. Exactness here means satisfying the
fixed distance criterion, not producing a unique graph topology when ties are
present.

At the common representative cell, the FastGraph axis control attains recall
$=1.0$, while neither approximate backend reaches exact recall at its recorded
highest-quality setting. Table~\ref{tab:recall} reports means over the three
repetitions of that cell. We do not report a cross-grid mean because the
archived default-build CAGRA data cover $d=2$--$7$, whereas the GGNN and
FastGraph data extend through $d=10$.

\begin{table}[!htbp]
\centering
\caption{Distance-based recall at $d{=}3$, $N{=}5\times10^5$,
$K{=}40$. CAGRA uses its default graph build with
\texttt{itopk\_size}=512; GGNN uses $\tau_{\text{query}}=0.8$.
Values are means over three repetitions.}
\label{tab:recall}
\small
\begin{tabular}{lc}
\toprule
Method & distance-based recall \\
\midrule
FastGraph axis control (exact) & $1.000$ \\
FAISS-GPU \texttt{IndexFlatL2} (exact) & $1.000$ \\
CAGRA, default build (approximate) & $0.818$ \\
GGNN (approximate) & $0.410$ \\
\bottomrule
\end{tabular}
\end{table}

These values establish the quality gap only in the recorded diagnostic
configurations. They do not measure recall for the headline timing campaign.

The PCA path is validated separately with the released
\texttt{binned\_select\_knn\_pca} entry point. Dedicated GPU tests compare
the complete sorted distance multiset returned for every query with an
element-wise FP32 brute-force reference on seeded Gaussian inputs at $d\in\{4,5,8,10\}$,
$K\in\{16,40,128\}$, and $N$ up to 8,000. They also exercise random PCA
subsampling and a three-segment \texttt{row\_splits} input. All tested
indices are valid, unique, and segment-local; returned distances match
recomputed distances, and every distance multiset matches brute force within
the test tolerances. Returned versus recomputed squared distances use
\texttt{atol}=$10^{-6}$ and \texttt{rtol}=$10^{-5}$; sorted distance
multisets versus brute force use \texttt{atol}=$10^{-6}$ and
\texttt{rtol}=$10^{-3}$. Each comparison requires
$|a-b|\le\texttt{atol}+\texttt{rtol}|b|$, with $b$ the reference value.
These finite tests validate the exercised configurations, not all FP32 inputs.

\begin{figure}[!htbp]
\centering
  \includegraphics[width=0.68\textwidth]{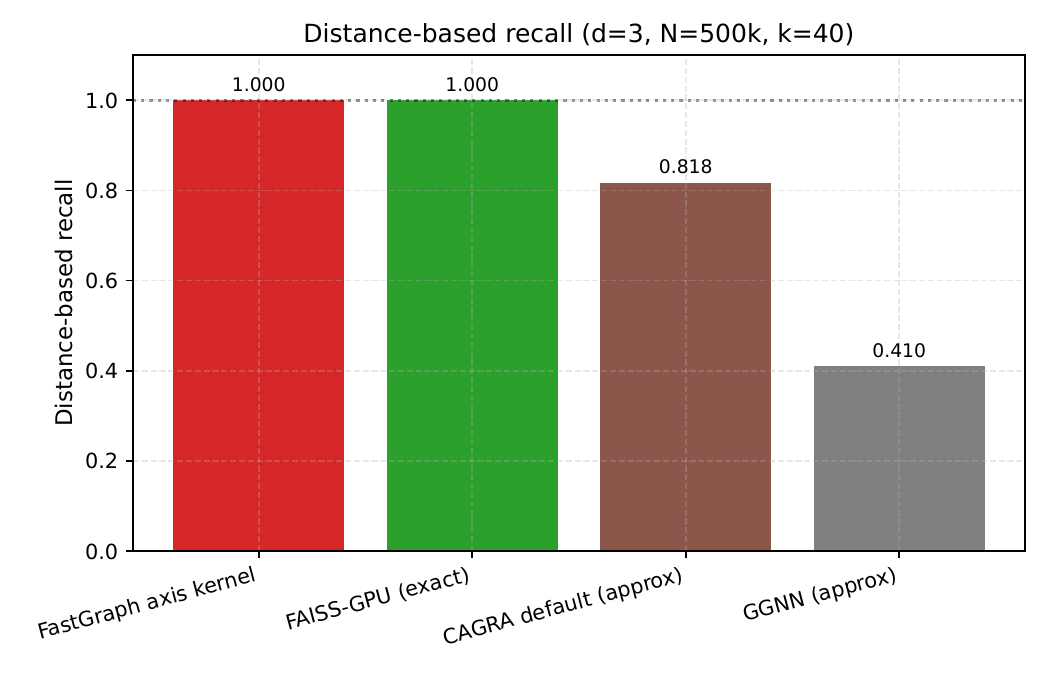}
  \caption{Distance-based recall at the representative HGCAL diagnostic cell
  $d{=}3$, $N{=}5\times10^5$, $K{=}40$. FastGraph and FAISS-GPU satisfy their
  exact distance conventions; the approximate results use default-build CAGRA
  with \texttt{itopk\_size}=512 and GGNN with
  $\tau_{\text{query}}=0.8$. This figure reports recall only.}
  \label{fig:recall_eval}
\end{figure}

\FloatBarrier
\subsection{Dynamic graph construction in GNNs}

GravNet-style layers~\cite{Qasim2019GravNet} construct a fresh kNN
graph in a learned latent space at every forward pass, so the
kNN substrate is on the gradient path of every training step.
FastGraph integrates into this setting through \texttt{GravNetOp}, a
PyTorch module that wraps coordinate transformation, kNN graph
construction (via the opt-in \texttt{use\_pca=True} PCA-subspace path of
Section~\ref{sec:method}), and message passing into a single
autograd-compatible operation. Gradients flow through the selected
distances and the input coordinates; the discrete neighbor indices
and the per-call projection $V_k$ are held fixed across each
forward--backward pair so that graph selection is consistent within the
pair; atomic gradient accumulation is not guaranteed to be bitwise
deterministic. Object condensation~\cite{Kieseler2020ObjectCondensation},
which is the loss used together with GravNet for HGCAL
reconstruction, is supported through a separate GPU-resident library
helper.

\section{Limitations}
\label{sec:limitations}

\paragraph*{Data-adaptive binning subspace}
Both FastGraph and its axis-aligned ablation bound the grid at
$n_{\text{bins}}^{k}$ for a small binning dimension $k$, independent of input
dimension $d$. FastGraph uses a data-adaptive orthonormal basis rather than
tying the grid to a fixed set of input columns. This removes
dependence on column order for a fixed feature matrix, but PCA is not
invariant to rescaling individual features: their units and preprocessing
change the covariance and therefore the selected subspace. The method
still requires a small $k$ to keep the grid tractable.

\paragraph*{Data distribution still matters}
The PCA contribution's effectiveness is a function of how much
pairwise-distance variance the leading components actually capture.
On the anisotropic detector-feature matrix used here, FastGraph's
PCA-subspace binning is substantially
faster than axis-aligned cell lists (Section~\ref{sec:mbd-ablation},
geomean speedups in Section~\ref{sec:headline}). On isotropic data, the
leading subspace is an arbitrary orthonormal $k$-dimensional projection and FastGraph
degrades gracefully toward ordinary cell-list behavior; in that regime
FastGraph outperforms dense GPU brute force
only while the cell-list pruning advantage outweighs the constant
factor of an optimized BLAS \texttt{gemm}. Section~\ref{sec:synthetic}
documents this on Gaussian $\mathcal{N}(0, I_d)$ inputs and is the
reason we do not claim FastGraph is the fastest GPU exact kNN on
arbitrary distributions; we claim it is the fastest among the tested
exact GPU baselines on HGCAL-like workloads where the leading
principal components carry real structural information. Whether trained
learned-latent embeddings exhibit similarly favorable spectra remains to be
tested.

\paragraph*{Remaining limitations}
PCA estimation adds a sampled low-rank fit and an $O(Ndk)$ projection
that are amortized over the kNN query in the regimes we benchmark, but
at very small $N$ the
projection cost becomes a non-negligible fraction of the total
wall-clock; for tiny per-event point counts an axis-aligned variant
remains preferable. The current implementation targets a single
GPU; multi-GPU or distributed kNN graph construction is not
supported. The discrete neighbor indices are not differentiable;
FastGraph's autograd contract is over the continuous distances and
coordinates only, matching every other hard-selection kNN layer.
Approximate methods can be faster in some cells: GGNN
wins on raw wall-clock for $d{=}2$--$6$, whereas FastGraph is faster than
GGNN for $d{=}7$--$10$ and faster than CAGRA throughout the headline sweep.
Recall at these timed operating points remains unmeasured.
The appropriate backend therefore depends on dimension, required graph
quality, and end-to-end cost.

The detector-derived timing study is a kernel stress test rather than a
deployment benchmark: it concatenates recHits and passes each slice as a
single segment. Event-preserving row splits can change both the problem
size and the spatial density seen by a cell list. Likewise, the benchmark
features are detector-derived inputs, not embeddings sampled from a
trained model. Event-batched timings and learned-latent coordinates are
therefore required before extrapolating the reported speedups to
end-to-end training throughput. We defer those model-level measurements
to future work and make no end-to-end throughput claim here. Extending
the evaluated input dimension
beyond $d{=}10$ is secondary to those two validation steps;
higher-dimensional feature-space graph layers remain outside the present
claim. A projected-space LBVH is another possible extension when uniform
bins cease to prune effectively.

\paragraph*{Why exactness matters for the downstream GNN}
The case for an exact $k$NN backend over a recall-tuned approximate
one is not a raw-recall argument but a sensitivity-of-downstream
argument. Adversarial-robustness studies of graph neural networks
show that node predictions can shift substantially under small,
structure-preserving edge perturbations to the input
graph~\cite{zugner2018adversarial}; the same literature also
documents that systematic reweighting of the neighborhood---e.g.,
replacing direct neighbors with a diffusion-smoothed neighbor
set---measurably alters downstream accuracy~\cite{klicpera2019diffusion}.
The neighbor-set mismatches injected by an approximate $k$NN are
analogous edge perturbations and depend on build-side index
hyperparameters; their event-level distribution and downstream effect
have not been measured here.
We do not claim the HGCAL reconstruction targets are unusually
sensitive to approximate-$k$NN edge noise---that would require a
dedicated model study beyond the scope of this paper---but the more
conservative choice for
a physics application that demands reproducible, systematics-controlled
inference is the exact graph, particularly when an exact backend is
competitive on wall-clock in the relevant operating regime
(Section~\ref{sec:headline}).

\paragraph*{CAGRA on heterogeneous data}
The default CAGRA IVF-PQ build path fails in the 5\,M-point timing campaign at $d \ge 5$
(Section~\ref{sec:expsetup}); the NN-Descent fallback succeeds
across the full dimensional sweep but is \spdHGCALcagradeight\
slower than FastGraph at the $d{=}8$ reference cell. We use the
NN-Descent numbers for the headline timing campaign; the archived recall
diagnostic is explicitly labeled as default-build CAGRA. This is an observed compatibility issue
for the tested dataset and software configuration, not evidence of a
generic CAGRA limitation or a diagnosis of the internal failure cause.

\section{Data and Code Availability Statement}
\begingroup
\sloppy
The FastGraph implementation described in this paper is available at
\url{https://github.com/AgarwalAarush/FastGraphCompute}, as a fork of
the upstream \texttt{FastGraphCompute} library.
The immutable \texttt{pca-fgc-paper-v1.3.4} release tags pin the exact
FastGraph, benchmark-harness, plotting, and CLOVER-fork revisions used
for the reported experiments; the release is built against the CUDA
12.1 toolchain and reports FastGraphCompute package version \texttt{1.2.0}.
The evaluated and recommended kernel surface covers binning subspace size
$k \in \{2, 3, 4, 5\}$; the tag also contains experimental $k=6,7$
dispatches, which are excluded from the headline configuration. The benchmarking harness and runner scripts
used to produce the figures live in a companion repository,
\url{https://github.com/AgarwalAarush/fgc-performance}. The CLOVER
head-to-head used a fork of
\url{https://github.com/ampslab/clover-knn} at
\url{https://github.com/AgarwalAarush/clover-knn} with local source
modifications (\texttt{k\_values} aligned to the paper's $k{=}40$, the
hardcoded synthetic-N template chain in \texttt{main()} disabled in favor
of the mesh path, and the build retargeted to \texttt{sm\_80} for A100).
The revised manuscript source is maintained at
\url{https://github.com/AgarwalAarush/fastgraph-paper}.
\endgroup

The detector-derived stress tests use CMS HGCAL simulation data (recHit
feature vectors) accessed via an internal collaboration dataset of
\hgcalRecHits\ calorimeter hits with up to ten stored numerical
features per hit; their names and units are not included in the available
artifact. The benchmark concatenates events and does not release
event boundaries or learned model embeddings. The data itself is internal
to the CMS collaboration;
synthetic Gaussian and uniform datasets used in
Sections~\ref{sec:synthetic} and~\ref{sec:clover} are generated on the
fly from fixed random seeds and are fully reproducible from the released
code. CMS collaboration members can reproduce the HGCAL results
directly; external parties should contact the corresponding author.

\section{Conclusion}

We have presented FastGraph, an exact $k$-nearest-neighbor graph-construction
primitive for GPU geometric deep learning. FastGraph uses a data-adaptive
PCA subspace for cell-list pruning, while a contraction argument
($V_k V_k^\top \preceq I_d$) preserves exactness in the original
input space. On the detector-derived recHit stress test,
across $d{=}2$--$10$ at the headline single-segment $N{=}5$\,M,
$k{=}40$ operating
point, FastGraph is the fastest among the tested dimension-general
\emph{exact} GPU
$k$NN baselines at every tested dimension, with speedups up to
$41\times$ over
FAISS-GPU \texttt{IndexFlatL2}, $19\times$ over cuVS brute force
(the strongest exact GPU baseline), and $17\times$ over the
approximate CAGRA-NN-Descent build---the CAGRA path that completed the
full $d{=}2$--$10$ sweep in our harness. The ablation
in Section~\ref{sec:mbd-ablation} isolates the PCA rotation rather
than incidental implementation differences as the primary
contribution: comparing each variant at its own optimum gives a
within-paper ratio of \fgcVanillaBestPCAfactor\ between PCA-subspace
and axis-aligned binning at $d{=}8$, $N{=}5$\,M, $k{=}40$.

This paper does not claim that FastGraph is the fastest GPU $k$NN on
every distribution or every dimensionality. Two boundaries of the
claim are documented. First, the method targets the $d{=}4$--$10$
low-to-moderate-dimensional regime around GravNet's learned space, but the
current timing data are detector-derived stress-test features rather than
learned embeddings; wider feature-space graph layers and trained-model
coordinates require separate evaluation. The 3D-only spatial-acceleration methods (CLOVER,
LBVH, RT-cores) are unavailable in the $d \ge 4$ regime by
construction, and at the largest $d{=}3$ HGCAL head-to-head point
($N{=}5$\,M, $k{=}40$), the CLOVER spatio-graph index is about
$100\times$ faster than FastGraph (Section~\ref{sec:clover}). In
that case FastGraph is the ordinary 3D axis-aligned cell list, not
the PCA-subspace path. The $d{=}3$ detector-data case does not
overlap with the regime FastGraph is designed for, but it delineates
the boundary of the claim. Second, on \emph{isotropic} data the PCA
projection has no structure to exploit and FastGraph degrades toward
ordinary cell-list behavior; for that case the relative ordering
against dense BLAS brute force is determined by the cell-list
pruning advantage alone, not by the PCA contribution. The detector-feature
gains are consistent with an anisotropic-data advantage, subject to the
feature-scale and feature-prefix controls identified above, and PCA-subspace binning
should be a generally useful substrate for dynamic graph layers in
scientific GNNs whose latent spaces have similar spectral structure.

\section*{Declaration of Generative AI and AI-Assisted Technologies in the Manuscript Preparation Process}
During the preparation of this work, the authors used OpenAI Codex for
language editing, software review, and reproducibility checks. The authors
reviewed and edited the output as needed and take full responsibility for the
content of the published article.

\bibliographystyle{elsarticle-num}
\bibliography{references}

\end{document}